\documentclass[aip,jcp,preprint,showkeys,superscriptaddress,floatfix,a4paper]{revtex4-1}
\usepackage[english]{babel}
\usepackage{graphicx}
\usepackage{amssymb}
\usepackage{amsmath}
\usepackage{amsfonts}
\usepackage{color}
\usepackage{soul}
\usepackage[update,prepend]{epstopdf}

\begin{document}

\title{On the applicability of density dependent effective interactions in cluster-forming systems}

\author{Marta Montes-Saralegui}
\affiliation{Institute for Theoretical Physics and Center for Computational Material Science (CMS), 
Technische Universit{\"a}t Wien, Wiedner Hauptstra{\ss}e 8-10, A-1040 Wien, Austria}

\author{Gerhard Kahl} 
\affiliation{Institute for Theoretical Physics and Center for Computational Material Science (CMS), 
Technische Universit{\"a}t Wien, Wiedner Hauptstra{\ss}e 8-10, A-1040 Wien, Austria}
\email{gerhard.kahl@tuwien.ac.at}

\author{Arash Nikoubashman}
\affiliation{Institute of Physics, Johannes Gutenberg University Mainz,
Staudingerweg 7, 55128 Mainz, Germany}
\email{anikouba@uni-mainz.de}

\date{\today}

\begin{abstract}
We systematically studied the validity and transferability of effective,
coarse-grained, pair potentials in ultrasoft colloidal systems. We focused 
on amphiphilic dendrimers, macromolecules which can aggregate into clusters 
of overlapping particles to minimize the contact area with the surrounding 
(implicit) solvent. Simulations were performed for both the monomeric and 
coarse-grained model in the liquid phase at densities ranging from infinite 
dilution up to values close to the freezing point. For every state point, each 
macromolecule was mapped onto a single interaction site and the effective pair 
potential was computed using a coarse-graining technique based on force-matching. 
We found excellent agreement between the spatial dendrimer distributions obtained from 
the coarse-grained and microscopically detailed simulations at low densities,
where the macromolecules were distributed homogeneously in the system. However, 
the agreement deteriorated significantly when the density was increased further and 
the cluster occupation became more polydisperse. Under these conditions, 
the effective pair potential of the coarse-grained model can no longer be computed 
by averaging over the whole system, but the local density needs to 
be taken into account instead.
\end{abstract}
  
\maketitle

\section{Introduction}
\label{sec:introduction}
Amphiphiles are chemical compounds consisting of both solvophilic and
solvophobic blocks. When the solute concentration surpasses a certain
threshold, these particles spontaneously self-assemble into micellar 
aggregates to minimize the interface between the solvophobic block and the 
surrounding solvent. The size and shape of the self-assembled superstructures 
depends mainly on the microscopic properties of the amphiphiles, allowing 
for, {\it e.g.} spherical, cylindrical and lamellar aggregates.\cite{israelachvilli:book:1992}
This peculiar ability makes amphiphilic molecules indispensable for a wide variety of 
applications,\cite{kataoka:advdrug:2001, patist:collsci:2002, schramm:annrep:2003} 
for example as cleaning agents or emulsifiers in the cosmetic and food industry.

In this work we focus on the self-assembly behavior of amphiphilic dendrimers, 
which consist of a solvophobic core and a solvophilic shell. Recently, these 
macromolecules have gathered an increasing amount of attention, due to their propensity 
to form long-lived colloidal crystals, where each lattice site is populated by an 
aggregate.\cite{mladek:prl:2006, *mladek:prl:2006B, mladek:prl:2008, moreno:prl:2007, 
zhang:prl:10, camargo:jstat:2010, wilding:jcp:2014, lenz:jpcb:2011, lenz:prl:2012, 
coslovich:sm:2011, montes:jpcm:2013, montes:jcp:2014} Simulation 
studies of these so-called cluster crystals have revealed a wide range 
of peculiar static and dynamic properties, which sets them apart
from conventional single occupancy crystals. For example, the lattice
constant of the cluster crystals is density-independent and therefore
external pressure does not lead to a compression of the lattice, but
rather to an increase of the occupation number.\cite{montes:jcp:2014}
The dynamics of this process are characterized by activated 
hopping of the constituent particles and merging of neighboring lattice 
sites.\cite{moreno:prl:2007, camargo:jstat:2010, coslovich:sm:2011, montes:jpcm:2013, 
montes:jcp:2014} Furthermore, reentrant melting and isostructural phase transitions
have been reported for this class of amphiphiles.\cite{zhang:prl:10, 
wilding:jcp:2014}

Due to computational limitations, the majority of previous simulation
studies relied on coarse-grained (CG) models, where the macromolecular 
amphiphiles were modeled as single interaction sites. The corresponding
effective pair interactions were usually obtained in the limit of infinite dilution,
and have then been employed to calculate system properties at considerably
higher densities.\cite{mladek:prl:2008, zhang:prl:10} However, this
strategy might lead to an inaccurate representation of the original microscopic model, 
since the transferability of the effective pair potentials from infinite dilution
to finite densities is {\it a priori} not obvious.\cite{louis:jpcm:2002} 
For instance, it has been demonstrated for homopolymer systems that additional 
corrections are necessary to provide a faithful CG representation of the 
microscopically resolved (MR) systems at finite densities.\cite{bolhuis:jcp:2001}

To shed more light on this question, we performed a systematic analysis
of amphiphilic dendrimers in the liquid phase at densities ranging
from infinite dilution up to close to the freezing point. We computed
the effective potentials at each state point using a force-matching CG 
algorithm.\cite{izvekov:jpcb:2005, izvekov:jcp:2005, noid:jcp:2008, izvekov:jcp:2010} 
We then compared the emerging structures in the MR and CG simulations using the effective
(pair) potentials obtained at infinite dilution, $\Phi_{\rm eff}^0$, and at the
corresponding density, $\Phi_{\rm eff}$. We discovered that the CG simulations 
with $\Phi_{\rm eff}^0$ failed at reproducing the MR structures even at the lowest 
investigated densities, and significantly underestimated the freezing density.
The CG simulations with $\Phi_{\rm eff}$ exhibited good agreement with the
microscopic reference simulations at low densities, where the system
predominantly consisted of isolated amphiphiles and small clusters. 
However, the agreement deteriorated rapidly as the density was increased 
further. 

We did not expect such a significant deviation between the two representations, because 
the effective potentials employed in the CG simulations were calculated from 
MR simulations conducted at exactly the {\it same} density and temperature. This 
discrepancy can be rationalized by considering the spatial density variations 
in the system, which became more pronounced as the amphiphiles started to aggregate
into clusters of heterogeneous size. These inhomogeneities make it impossible to 
represent {\it all} relevant interactions by a {\it single} effective pair potential. 
To validate our hypothesis, we systematically calculated the effective potential between 
aggregates of different sizes, and found a highly non-linear relationship between the 
effective interaction and the aggregation number.

The rest of this manuscript is organized as follows. In 
Section~\ref{sec:systems} we introduce the investigated 
model systems and simulation method. The employed CG 
algorithm is presented in Section~\ref{sec:CG}. We discuss
our findings in Section~\ref{sec:results}, where we 
systematically compare our results from the microscopically
resolved simulations with the coarse-grained ones. Finally, we 
summarize the findings and draw our conclusions in Section~\ref{sec:conclusion_outlook}.

\section{Models and Methods}
\label{sec:methods}

\subsection{Systems and simulation method}
\label{sec:systems}
\begin{figure}[htbp]
\includegraphics[width=5cm]{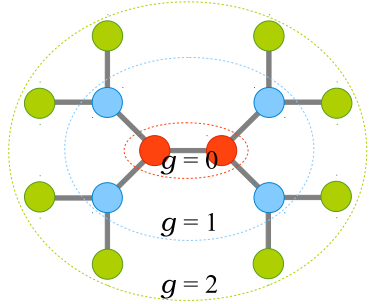}
\caption{Schematic representation of a dendrimer with parameters
  $f=3$, $p=1$ and $G=2$ (see text). Each shell of monomers is
  specified by its respective generation index $g$. Inner core
  monomers are colored in red and blue, while the outer shell
  monomers are shown in green.}
\label{fig:scheme}
\end{figure}

Dendrimers are characterized by a highly branched architecture, which is
specified by the functionality, $f$, the spacer length, $p$, and the total
number of generations, $G$. Dendrimers are grown by attaching $(f-1)$ 
chains with $p$ monomers to two bonded central monomers (generation index $g=0$). 
This process was repeated $G$ times, resulting in a self-similar structure. 
Figure \ref{fig:scheme} shows a schematic representation of the resulting 
dendritic macromolecule.

All MR simulations were conducted using a bead-spring model, where the constituent
monomers were represented by spherical beads with diameter $\sigma$ that were tethered 
through massless springs. Three different cases were considered in this work: one 
non-clustering dendrimer model under good solvent conditions,\cite{goetze:jcp:2004} 
and two different amphiphilic dendrimer models,\cite{mladek:prl:2008, lenz:prl:2012} 
denoted henceforward by I and II.

In all three cases the dendrimers are specified by $f=3$, $p=1$, and $G=2$, consisting 
thus of $n_{\rm m}=14$ monomers. Furthermore, the mass of the beads was set to unity, $m=1$. 
Covalent bonds were mimicked {\it via} the finite extensible non-linear elastic (FENE) 
potential:\cite{bishop:jcp:1979, welch:mm:1998}
\begin{align}
	\beta \Phi^{\rm FENE}(r_{ij}) = 
	- \kappa_{ij} R_{ij}^2\ln \left[1 - \left(\frac{r_{ij}-l_{ij}}{R_{ij}} \right)^2\right],
\label{eq:fene}
\end{align} 
with the reciprocal thermodynamic temperature of the system $\beta = 1/(k_{\rm B}T)$
and interparticle distance $r_{ij} = \left|\mathbf{r}_j-\mathbf{r}_i\right|$. 
The spring constant was controlled through the parameter $\kappa_{ij}$, whose 
magnitude depended on the identity of particles $i$ and $j$. The minimum and maximum
extension of the bond, $l_{ij}^{\rm min}$ and $l_{ij}^{\rm max}$ respectively, was set through 
the parameters:
\begin{align*}
	R_{ij} = (l_{ij}^{\rm max} - l_{ij}) ~~~~ {\rm and} ~~~~ l_{ij} =(l_{ij}^{\rm max} + l_{ij}^{\rm min})/2 .
\end{align*} 

For the non-clustering dendrimers at good solvent conditions, all the beads had the same identity. 
Excluded volume interactions were modeled using the purely repulsive Weeks-Chandler-Anderson 
(WCA) potential:\cite{weeks:jcp:1971}
\begin{align}
	\beta \Phi^{\rm WCA}(r_{ij})=
	\begin{cases}
	4\varepsilon\left[\left(\frac{\sigma}{r_{ij}}\right)^{12} -
	\left(\frac{\sigma}{r_{ij}}\right)^6\right] + \varepsilon & r_{ij} \leq 2^{1/6}\,\sigma \\
	0 & {\rm otherwise},
	\end{cases}
\label{eq:wca}
\end{align} 
where the interaction strength is quantified through $\varepsilon$.

In the case of the amphiphilic dendrimers, we distinguished between solvophobic core (C)
and solovophilic shell (S) particles. Here, the bead-bead interactions were modeled {\it via} 
the Morse potential:
\begin{align}
	\beta \Phi^{\rm Morse}(r_{ij})=
	\varepsilon_{ij} 
	\left\{\left[e^{-\alpha_{ij}(r_{ij}-\sigma_{ij})}-1\right]^2 -1 \right\}, 
\label{eq:morse}
\end{align}
with $\sigma_{ij} = (\sigma_i + \sigma_j)/2$. The parameters $\varepsilon_{ij}$ 
and $\alpha_{ij}$ controlled the strength and range of the interaction, respectively.

We used the interaction parameters from Ref.~\citenum{goetze:jcp:2004} for the
non-clustering dendrimers at good solvent conditions and the interaction parameters 
from Ref.~\citenum{mladek:prl:2008} (model I) and Ref.~\citenum{lenz:prl:2012} (model II)
for the amphiphilic dendrimers. The specific values are summarized in Table~\ref{tab:phiparam}.
If not stated otherwise explicitly, we used in all our MR simulations
$\sigma=\sigma_{\rm CC}$ as our unit of length, and $m$ as our unit of mass. 
From these units the intrinsic time unit of the MD simulations can be derived as $\tau=\sigma \sqrt{\beta m}$. 
Densities are defined as $\rho = N/R_{\rm g}$, where $N$ is the total number of dendrimers
in the system and $R_{\rm g}$ is the radius of gyration of the dendrimers [see Eq.~(\ref{eq:rg})]. 
The overlap density of  the polymer solution, $\rho_{\rm overlap}$, is defined as
$\frac{4}{3}\pi(\frac{3}{2}R_{\rm g})^3 \rho_{\rm overlap} = 1$.

\begin{table*}[htbp]
\begin{center}
\begin{tabular}{c|ccccc|ccccc}
\hline
\hline
\hspace*{0.0cm} Model \hspace*{0.0cm} & \hspace*{0.0cm} Type \hspace*{0.0cm} & 
\hspace*{0.0cm} Interaction \hspace*{0.0cm} & \hspace*{0.15cm} $\epsilon_{ij}$ \hspace*{0.15cm} & 
\hspace*{0.15cm} $\sigma_{ij}$ \hspace*{0.15cm} & \hspace*{0.15cm} $\alpha_{ij}$ \hspace*{0.15cm} & 
\hspace*{0.0cm} Type \hspace*{0.0cm} & \hspace*{0.0cm} Interaction \hspace*{0.0cm} &
\hspace*{0.15cm} $\kappa_{ij}$ \hspace*{0.15cm} & \hspace*{0.15cm} $l_{ij}$ \hspace*{0.15cm} & 
\hspace*{0.15cm} $R_{ij}$ \hspace*{0.15cm} \\
\hline
NC &  & WCA & 1 & 1 & & & FENE & 0.5 & 0 & 10\\
\hline
& CC & Morse & 0.714 & 1 & 6.4 & CC & FENE & 40 & 1.875 & 0.375\\
I & CS & Morse & 0.014 & 1.25 & 19.2 & CS & FENE & 20 & 3.75 & 0.75\\
&SS & Morse & 0.014 & 1.5 & 19.2 & & & &\\
\hline
&CC & Morse & 0.714 & 1 & 1.8 & CC ($g=0$) & FENE & 60 & 3.1875 & 0.6375\\
II & CS & Morse & 0.01785 & 1.75 & 6.0 & CC ($g\neq0$) & FENE & 60 & 1.875 & 0.375\\
&SS & Morse & 0.01785 & 2.5 & 6.0 & CS & FENE & 30 & 3.5625 & 0.7125\\
\hline
\hline
\end{tabular} 
\end{center}
\caption{Numerical parameters for the non-bonded (left) and bonded (right)
interactions used in the MR simulations of non-clustering (NC) and amphiphilic
dendrimers (model I and II). The abbreviations C and S refer to the different 
types of monomers involved in the respective interactions.}
\label{tab:phiparam}
\end{table*} 

Molecular dynamics simulations were conducted in the $NVT$ ensemble using the LAMMPS 
simulation package.\cite{lammps} The velocity Verlet algorithm\cite{swope:jcp:1982, 
allen:book:1989, frenkel:book:2001} was employed for integrating the equations of motion, 
with a timestep of $\Delta t=5 \times 10^{-4}$ for the MR simulations and $\Delta t=5 \times 10^{-3}$ 
for the CG simulations. In the CG simulations we set the mass of the effective particles to 
unity, which introduced a factor of $\sqrt{n_{\rm m}}=\sqrt{14}$ between the time units in 
the MR and the CG picture.

The temperature was fixed to $T=1$ through a Nos{\'e}-Hoover thermostat.\cite{nose:jcp:1984, 
nose:molphys:1984, hoover:pra;1985, hoover:pra:1986} The central idea of this scheme is to 
couple the system to an (implicit) external heat reservoir through a fictitious spring, allowing 
for heating as well as for dissipation of excess heat. Here, the coupling strength can be tuned 
{\it via} the damping time of the spring, $t_{\rm d}$. On the one hand, too large values of 
$t_{\rm d}$ (loose coupling) may cause poor temperature control, whereas on the other hand, 
too small values (tight coupling) may cause high-frequency temperature oscillation. We found 
that $t_{\rm d}=0.09$ led to quick equilibration as well as good stability, and therefore 
used this value in all our simulations.

The initial configurations for the MR simulations were generated by 
growing each of the $N$ dendrimers along a self-avoiding random walk in a cubic simulation box.
The systems were initialized in a highly diluted state, where the individual dendrimers were
essentially isolated from each other. From these states, the final starting configurations were 
created by slowly compressing the simulation box until the desired density was reached. Once the 
starting configurations were produced, the systems were equilibrated 
until the potential energy did not change anymore. 

\subsection{Coarse-graining method}
\label{sec:CG}
In many situations, microscopically resolved simulations are computationally unfeasible
due to the vast number of interaction sites. Such microscopic simulations are further 
impeded by the relatively small timesteps, which are required for capturing the 
dynamics of the particles. Fortunately, the microscopic details of the individual 
macromolecules are often only of minor interest and can therefore be suitably traced out 
for the sake of computational efficiency. The acceleration achieved through such a 
coarse-graining is twofold: first, the number of interaction sites is reduced 
dramatically through this procedure, which facilitates the force calculation and the 
integration of the equations of motion. Second, CG models exhibit inherently faster 
dynamics compared to their MR counterparts, since the fast internal degrees of 
freedom have been integrated out.\cite{karimi:cpc:2012}

Various techniques have been developed to map the complex interactions of the MR system onto 
effective pair potentials, for example iterative Boltzmann inversion,\cite{lomba:pre:2003, 
reith:jcc:2003} force matching,\cite{izvekov:jpcb:2005, izvekov:jcp:2005, noid:jcp:2008, izvekov:jcp:2010} 
or physically informed {\it ad-hoc} models.\cite{pike:jcp:2009, mueller:langmuir:2014} 
The mapping from the MR to the CG picture is in general {\it not} unique but depends on which of the 
physical quantities from the original MR system should be conserved.\cite{louis:jpcm:2002} 
In this work, we employed the so-called multi-scale coarse graining (MSCG) method developed 
by Voth and co-workers.\cite{izvekov:jpcb:2005, izvekov:jcp:2005, noid:jcp:2008, izvekov:jcp:2010}

In what follows, we provide a short description of the employed CG approach. We
distinguish between quantities in the MR and CG pictures by denoting the corresponding
properties with lower- and upper-case symbols, respectively. Each dendrimer was mapped to a single 
interaction site, which was located at the center-of-mass (CM) of the original macromolecule:
\begin{align}
	\mathbf{r}_{{\rm CM}, I} = \mathbf{R}_I = 
	\frac{1}{n_{\rm m}} \sum_{i=1}^{n_{\rm m}}\mathbf{r}_i ,
\label{eq:com}
\end{align} 
where $\mathbf{r}_i$ denotes the position of monomer $i$ of the $I$-th
dendrimer (omitting in the following the index $I$ for clarity). The size of the
coarse-grained particle was defined through the dendrimer's radius of 
gyration:
\begin{align}
	R_{\rm g}^2 = \frac{1}{n_{\rm m}-1} \sum_{i=1}^{n_{\rm m}} \left(\mathbf{r}_i - \mathbf{r}_{{\rm CM}, I}\right)^2 .
	\label{eq:rg}
\end{align}

We assumed that the effective interaction between the CG particles depended
only on the interparticle distance and that all interactions were pairwise additive. 
Furthermore, we introduced a cutoff radius, $R_{\rm max}$, beyond which 
CG particles did not interact with each other. In the MSCG framework, the 
force acting on the effective particle $I$, ${\bf F}_I ({\bf R}^N)$, is 
given by:
\begin{align}
	\textbf{F}_I(\textbf{R}^N) = \sum_{J=1, J\neq I}^{N} f(R_{IJ})\hat{\textbf{R}}_{IJ}.
	\label{eq:cg_force}
\end{align}
Here, $\textbf{R}^N$ represents the entire set of the $\textbf{R}_I$
(with $I = 1, \dots, N$), $R_{IJ} = \left|\mathbf{R}_{IJ}\right| = 
\left| \mathbf{R}_J - \mathbf{R}_I\right|$ is the distance between particles 
$I$ and $J$, and $\hat{\textbf{R}}_{IJ}$ denotes the unit vector pointing 
along $\mathbf{R}_{IJ}$. 

The function $f(R)$ in Eq.~(\ref{eq:cg_force}) is non-zero in the range 
$0\leq R \leq R_{\rm max}$ and needs to be determined from the MR simulations. 
For the explicit evaluation of $f(R)$, we divided the interval $[0, R_{\rm max}]$ 
into $N_{\rm D}$ equally spaced sub-intervals and performed a piecewise 
decomposition of $f(R)$ into a sum over basis-functions, $f_d(R)$:

\begin{align}
	f(R) = \sum_{d=0}^{N_D} \phi_d f_d(R) ,
	\label{eq:cg_force2}
\end{align}
where the $f_d(R)$ are linear splines with yet unknown coefficients $\phi_d$.
The functional form of a spline in the interval $[R_{d-1}, R_{d+1}]$ is given by:

\begin{align}
f_{d}\left(R\right) \equiv
	\begin{cases}
		\frac{R-R_{d-1}}{R_{d}-R_{d-1}}&
		\text{if $R_{d-1}$ $<$ $R$ $\leqslant$ $R_{d}$}  \vspace{0.2cm} \\		
		\frac{R_{d+1}-R}{R_{d+1}-R_{d}} &
		\text{if $R_{d}$ $<$ $R$ $\leqslant$ $R_{d+1}$} \vspace{0.2cm} \\
		0  &\text{otherwise} .\\
	\end{cases}
\end{align}
The actual values of $N_D$ and $R_{\rm max}$ are system-specific and
will be reported in the corresponding subsections where the respective 
results are presented and discussed.

The effective potentials were then calculated by determining the coefficients
$\phi_d$ in Eq.~(\ref{eq:cg_force2}). Substituting $\textbf{R}^N$ and 
$\textbf{F}^N$ by $\textbf{r}^N_{\rm CM}$ and $\textbf{f}^N_{\rm CM}$ led 
to a set of $N$ linear equations of the $N_D$ parameters,
which were solved using the least-squares (LSQR) algorithm.\cite{paige:ACM:1982, 
paige:ACM:1982b} To improve sampling, the parameters were computed and
averaged from $n_t$ statistically independent configurations.
In practice, snapshots were taken every 500 to 5000 timesteps, and we
carefully checked that the solutions converged by continuously
increasing $n_{\rm t}$ until the results did not change anymore. We
found that 10000 - 25000 configurations were in general sufficient to
meet these requirements.

\section{Results}
\label{sec:results}

\subsection{Dendrimers in a good solvent}
In order to test our implementation of the MSCG algorithm, we first
studied a system of dendrimers in a good solvent. Under these conditions,
the macromolecules should not exhibit any clustering, but should be
distributed homogeneously in the system.
In Ref.~\citenum{goetze:jcp:2004}, G{\"o}tze {\it et al.} calculated
the effective potential of dendrimers ($G=4$, $p=1$, and $f=3$) in a
good solvent at $T=1$. In the limiting case of infinite dilution, {\it i.e.} 
for $\rho \to 0$, an effective potential $\Phi_{\rm eff}^0(r)$ with a Gaussian 
shape was obtained. The transferability of this model was tested by 
conducting additional MR simulations for densities up to the overlap density
and comparing the resulting pair correlation functions $g(r)$ with the ones
obtained from CG simulations using $\Phi_{\rm eff}^0(r)$.\cite{goetze:jpcm:2005}
Excellent agreement was observed for the entire density range with only 
slightly higher ordering observed in the MR simulations. This discrepancy
was attributed to the deformation of the individual dendrimers in
the MR simulations, which was not included in the employed CG model.

\begin{figure}[htbp]
	\begin{center}
	\includegraphics[width=8.6cm]{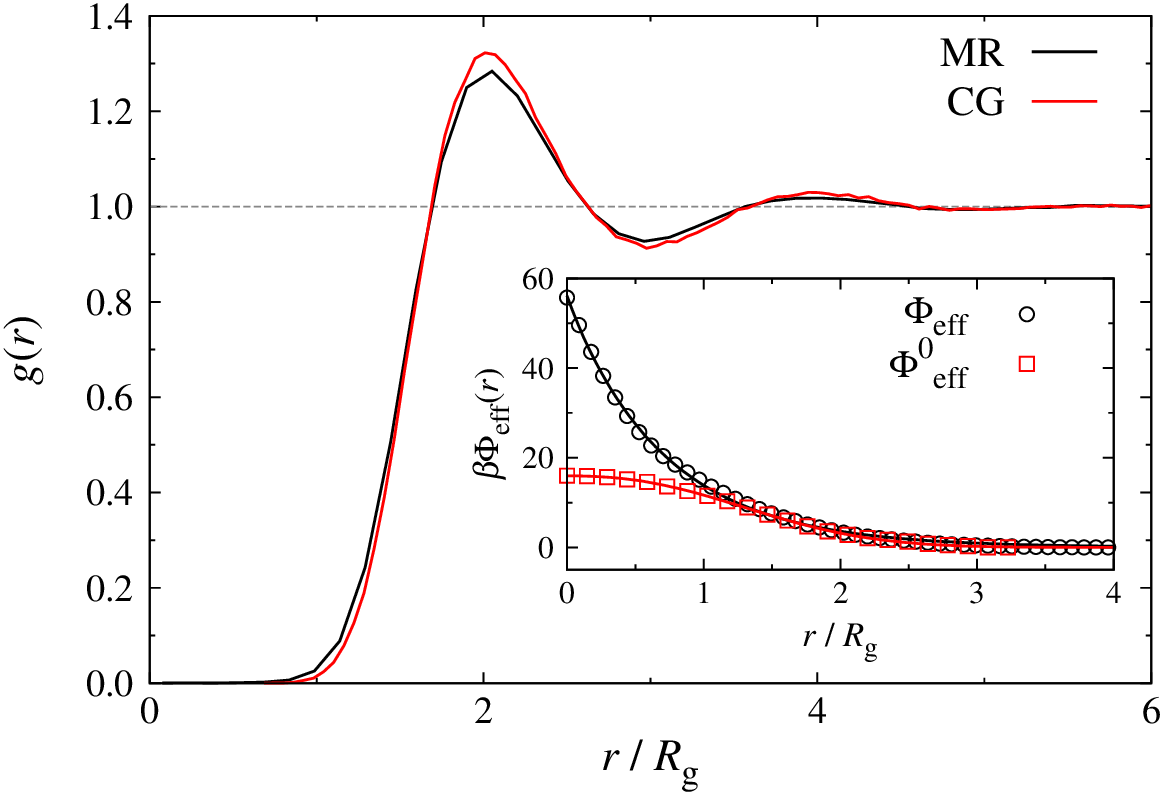} 
	\end{center}
	\caption{Comparison of the radial distribution function $g(r)$
	computed both in MR and in CG simulations (as labeled), carried out
	at $\rho=2\,\rho_{\rm overlap}$. The CG simulation has 
	been conducted with an effective pair potential, $\Phi_{\rm eff}(r)$, 
	computed at the same density as the MR simulation; $\Phi_{\rm eff}(r)$ 
	is shown in the inset together with its fit to $\Phi_{\rm GEM}^n(r)$ 
	with $n=0.97$.}
	\label{fig:athermal}
\end{figure}

In this contribution, we extended the density range to $\rho =
2\rho_{\rm overlap}$ using $N=500$ dendrimers. We measured a
radius of gyration of $R_{\rm g}=3.41$ for these macromolecules, and
computed the effective pair interaction using a set of $N_{\rm D}=8$
basis functions and a cutoff radius $R_{\rm max}=20$. The main panel
of Figure \ref{fig:athermal} shows a comparison of the CM $g(r)$, computed 
both in the MR and CG simulations. A remarkable agreement between the 
data is evident, confirming that CG simulations can be used even at 
densities significantly larger than $\rho_{\rm overlap}$.

The inset of Figure \ref{fig:athermal} shows the corresponding effective 
potential $\Phi_{\rm eff}(r)$ and it is well visible that it changed 
significantly compared to its form at infinite dilution, $\Phi_{\rm eff}^0(r)$;
as $\rho$ was increased beyond $\rho_{\rm overlap}$, the effective pair potential 
became significantly steeper at the origin. This effect can be attributed to the 
steric interactions between the monomers, which impeded the overlap of nearby 
dendrimers at high densities.

To obtain a functional form for $\Phi_{\rm eff}(r)$, we fitted the 
computed potential {\it via} the generalized exponential model of index 
$n$ (GEM-$n$):\cite{mladek:prl:2006, *mladek:prl:2006B}
\begin{align}
\Phi_{\rm GEM}^n(r) = 
\varepsilon_{\rm GEM} \exp\left[-\left(r/\sigma_{\rm GEM}\right)^n\right] ,
\label{eq:GEM}
\end{align}
where $\varepsilon_{\rm GEM}$ parameterizes the strength of the
potential, $\sigma_{\rm GEM}$ is the diameter of the effective particle, 
and $n$ controls the steepness of the shoulder. For $n=1$,
$\Phi_{\rm GEM}^{1}(r)$ decays exponentially, while this function
becomes a Gaussian for $n=2$. In the limit of $n \to \infty$,
$\Phi_{\rm GEM}^\infty(r)$ becomes a square shoulder potential. It has
been demonstrated in Ref.~\citenum{likos:pre:2001}, that particles
interacting \textit{via} a $\Phi_{\rm GEM}^n(r)$ potential exhibit
clustering if $n>2$.

We obtained an exponent of $n=0.97$ for the simulations conducted at 
$\rho = 2\,\rho_{\rm overlap}$, whereas $n=2$ was found for the situation
at infinite dilution. Both values are below the theoretically estimated 
threshold for clustering, which is in agreement with the expected behavior
for these macromolecules dispersed in a good solvent.

\subsection{Amphiphilic dendrimers - Model I}
\label{sec:model1}
In order to study the density dependence of the effective potentials
in {\it clustering} systems, we first simulated the amphiphilic dendrimers systems
investigated in Ref.~\citenum{mladek:prl:2008}. These dendrimers had
a functionality of $f=3$, a spacer length of $p=1$, and were terminated
in their growth after $G=2$ generations (see Figure~\ref{fig:scheme}). 
Mladek {\it et al.} calculated the zero-density effective potential for these
amphiphiles,\cite{mladek:prl:2008} and demonstrated that it fulfills
the clustering criterion derived in
Ref.~\citenum{likos:pre:2001}. Lenz {\it et al.} attempted to verify
the validity of the CG picture at finite densities by performing MR
simulations in the fluid state.\cite{lenz:jpcb:2011} They found
qualitative agreement between the pair correlation functions in the MR
and CG simulations (using the zero-density effective potential) at low
and intermediate densities.

In order to provide a more quantitative analysis, we systematically computed the 
effective potentials at five {\it finite} densities, {\it i.e.} $\rho = 0.38$, 
$0.52$, $0.65$, $0.82$, and $1.05$. This density range covers the state points 
investigated in Ref.~\citenum{lenz:jpcb:2011}. We simulated an ensemble of $N=2000$ 
dendrimers for all densities ($N=500$ for $\rho = 0.38$) to ensure proper sampling of 
the measured system properties.

\begin{figure}[htbp]
	\begin{center}
	\includegraphics[width=8.6cm]{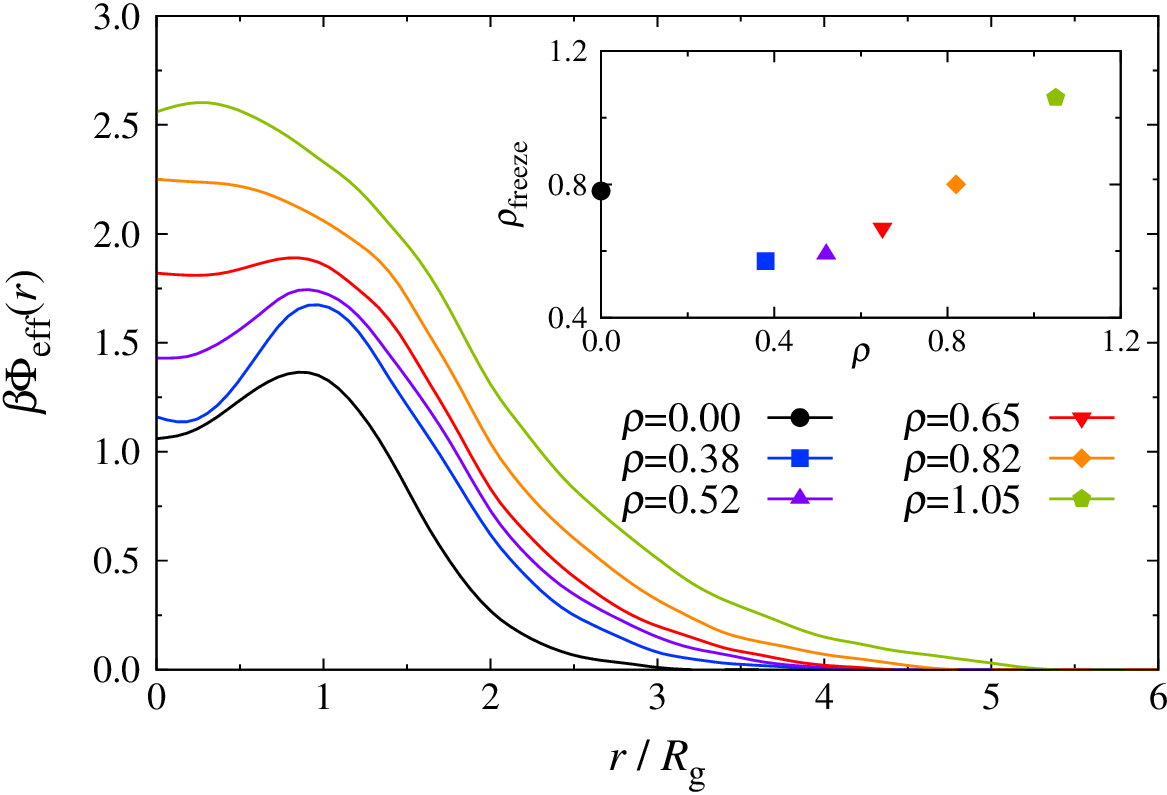} 
	\end{center}
	\caption{Effective potential $\Phi_{\rm eff}(r)$ of a system of amphiphilic dendrimers 
	(model I) computed at different densities (as labeled). The $x$-axis has been scaled by the 
	radius of gyration $R_{\rm g}$ at the respective density $\rho$. Inset: freezing density 
	calculated from Eq.~(\ref{eq:freeze}) using the effective pair potential computed at 
	the $\rho$-value specified in the $x$-axis.}
	\label{fig:potential_amphi1}
\end{figure}

The main panel of Figure~\ref{fig:potential_amphi1} shows all $\Phi_{\rm eff}(r)$, 
which were computed using the MSCG algorithm with a basis of dimension $N_{\rm D}=13$ 
and a cutoff radius $R_{\rm max}=20.0$. The effective potential $\Phi_{\rm eff}(r)$ 
between two isolated amphiphiles ($\rho \to 0$) has a local minimum at $r=0$, 
which corresponds to a configuration where the solvophobic cores of the two 
dendrimers overlap. The effective potential had its maximum at 
$r \approx R_{\rm g}$, {\it i.e.} when the solvophilic shell of a dendrimer 
penetrates the solvophobic core of another one, and {\it vice versa}.

Our analysis revealed a strong density dependence of $\Phi_{\rm eff}(r)$: 
the long-ranged repulsion of the potential increased monotonically with 
$\rho$, while the short-ranged attraction, characterized by 
$\Delta\Phi_{\rm eff} = \Phi_{\rm eff}(R_{\rm g}) - \Phi_{\rm eff}(0)$, 
gradually disappeared. However, we found an intermediate 
density range $0 < \rho < 0.52$, where $\Delta\Phi_{\rm eff}$ increased, 
indicating an enhanced affinity to form clusters. 

The state point at which the system freezes into a cluster crystal can be estimated 
accurately by minimizing the free energy of the crystal with respect to the cluster 
occupation number and lattice spacing.\cite{mladek:prl:2006, *mladek:prl:2006B,
likos:jcp:2007} Following the arguments brought forward in Ref.~\citenum{likos:jcp:2007},
the freezing density, $\rho_{\rm freeze}$, can be calculated directly from the effective 
potential $\Phi_{\rm eff}(r)$ through:
\begin{align}
	\frac{T_{\rm freeze}}{\rho_{\rm freeze}} \approx 1.393 \left| \tilde{\Phi}_{\rm eff}(k_{\rm min}) \right|
	\label{eq:freeze}
\end{align}
where $T_{\rm freeze}$ is the freezing temperature, $\tilde{\Phi}_{\rm eff}(k)$ is the
Fourier transformation of $\Phi_{\rm eff}(r)$, and $k_{\rm min}$ is the position of
the minimum of $\tilde{\Phi}_{\rm eff}(k)$. One peculiar property of Eq.~(\ref{eq:freeze})
is the constant ratio between $T_{\rm freeze}$ and $\rho_{\rm freeze}$, which leads to 
a straight freezing line in the phase diagram.\cite{likos:jcp:2007} 

We computed $\rho_{\rm freeze}$ for each $\Phi_{\rm eff}(r)$ at a fixed freezing temperature 
$T_{\rm freeze}=T=1$, and plotted the data in the inset of Figure~\ref{fig:potential_amphi1}. 
Here, we can see that $\rho_{\rm freeze}$ changed significantly as $\rho$ was increased: 
for $\rho \leq 0.65$, $\rho_{\rm freeze}$ decreased with respect to the freezing density in 
the zero-density limit, $\rho^0_{\rm freeze}$. In fact, $\rho_{\rm freeze}$ attained its 
minimum at $\rho=0.38$, the same density where $\Delta \Phi_{\rm eff}$ is maximized. 
These data suggest that, initially, clustering was enhanced by the presence 
of additional amphiphiles. At low densities, it was beneficial to place 
dendrimers on top of each other, since this strategy decreased the contact 
area of the solvophobic cores with the surrounding solvent. However, as the 
density was increased further, excluded volume effects made it increasingly 
difficult to place additional dendrimers into a cluster. This interpretation 
is supported by the slight swelling of the dendrimers, quantified {\it via}
the increase of $R_{\rm g}$ from $R_{\rm g} = 3.30$ to $R_{\rm g} = 3.36$ 
(measurement uncertainty $\pm 0.01$) as the density was increased. 

\begin{figure}[htbp]
	\begin{center}
	\includegraphics[width=8.6cm]{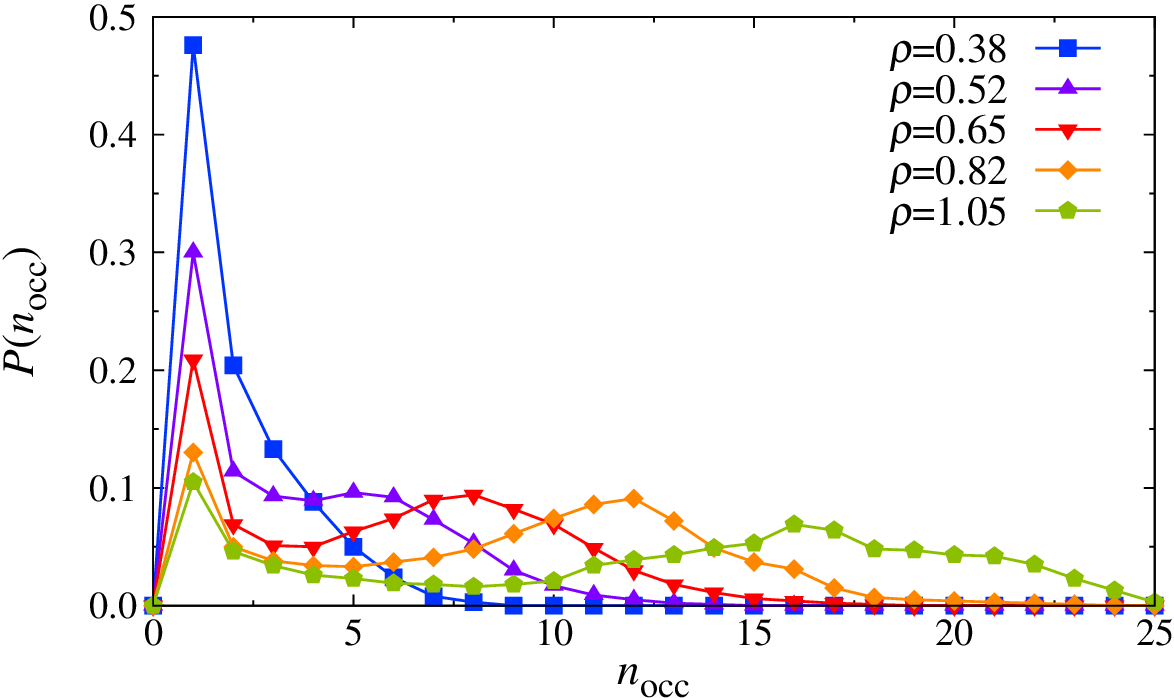} 
	\end{center}
	\caption{Cluster size distribution $P(n_{\rm occ})$ in the MR simulation of amphiphilic 
	dendrimers (model I) at different densities $\rho$ (as labeled).}
	\label{fig:cluster_amphi1}
\end{figure}

Figure \ref{fig:cluster_amphi1} shows the cluster size distribution in the 
MR simulations, obtained by applying a simple distance-based cluster analysis 
algorithm, where dendrimers with a CM separation of less than 
$r_{\rm c}=0.9\,R_{\rm g}$ were assigned to the same cluster. For $\rho=0.38$, 
approximately half of all dendrimers were isolated and $P(n_{\rm occ})$ 
decreased monotonically with $n_{\rm occ}$. For $\rho=0.52$, the number of 
isolated dendrimers decreased to $P(1) = 30\,\%$ and we found a local maximum 
of $P(n_{\rm occ})$ at $n_{\rm occ}=6$, indicating the onset of clustering. 
As $\rho$ was increased further, $P(1)$ kept decreasing while the local maximum 
of $P(n_{\rm occ})$ shifted towards higher values of $n_{\rm occ}$.

\begin{figure}[htbp]
	\begin{center}
	\includegraphics[width=8.6cm]{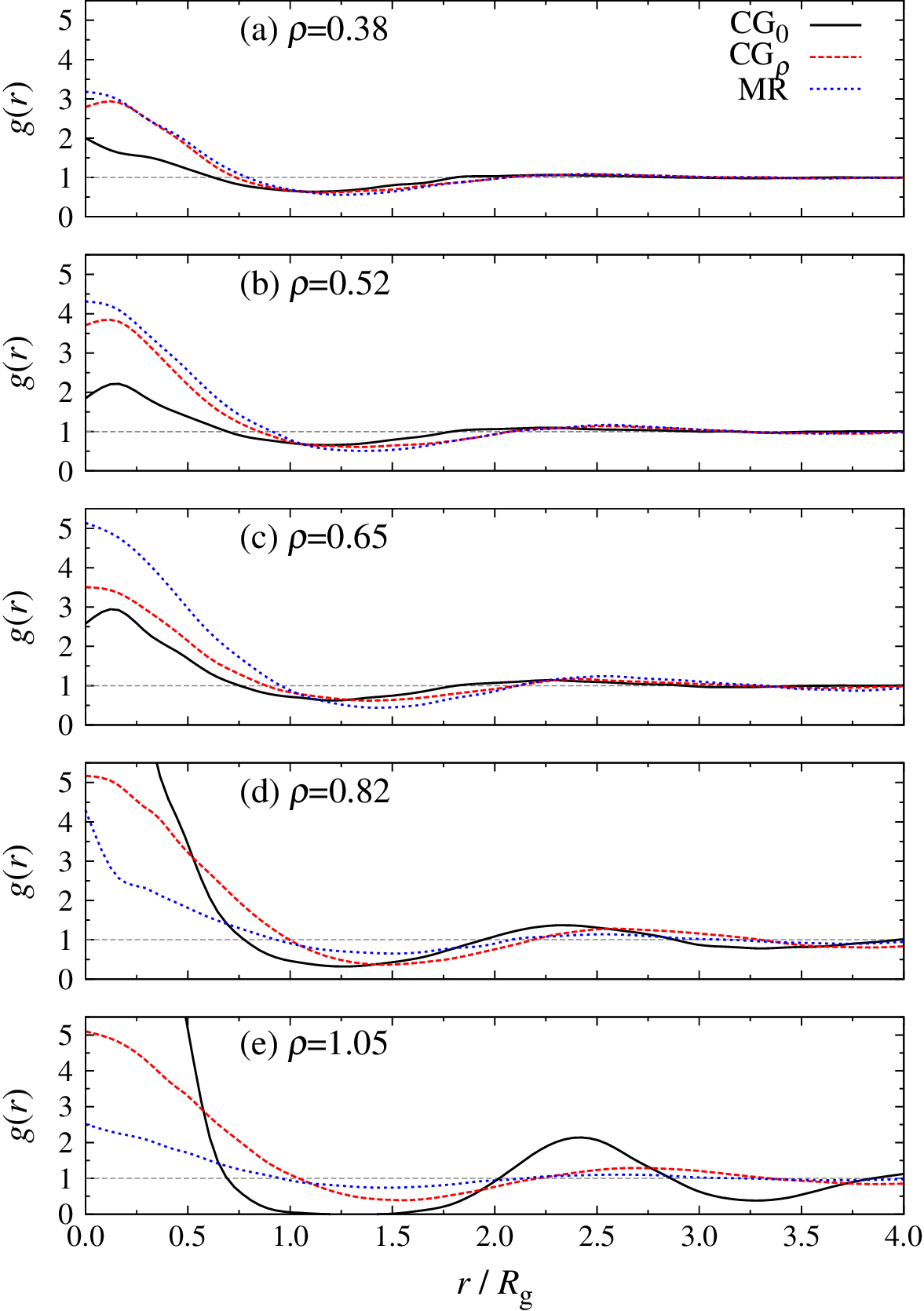} 
	\end{center}
	\caption{Comparison of the radial distribution function $g(r)$ computed 
	in a MR simulation (solid line), a CG simulation using the potential 
	computed at the corresponding density (dashed line) and a CG simulation 
	using the effective potential computed at infinite dilution (dash-dotted 
	line) in a system of amphiphilic dendrimers (model I).}
	\label{fig:rdf_amphi1}
\end{figure}

We then compared the resulting radial distribution functions $g(r)$ of 
the MR simulations with the ones from the CG simulations using 
$\Phi_{\rm eff}(r)$ and $\Phi_{\rm eff}^0(r)$. It is evident from 
Figure~\ref{fig:rdf_amphi1} that the CG simulations using $\Phi_{\rm eff}^0$ 
failed completely in reproducing the structures observed in the MR simulations
even at the lowest investigated density ($\rho = 0.38$). In contrast, the agreement 
between the MR simulations and the CG calculations with $\Phi_{\rm eff}(r)$ was 
significantly better up to $\rho \lesssim 0.52$, but then worsened rapidly 
as the density was further increased. 

\begin{figure}[htbp]
	\begin{center}
	\includegraphics[width=8.6cm]{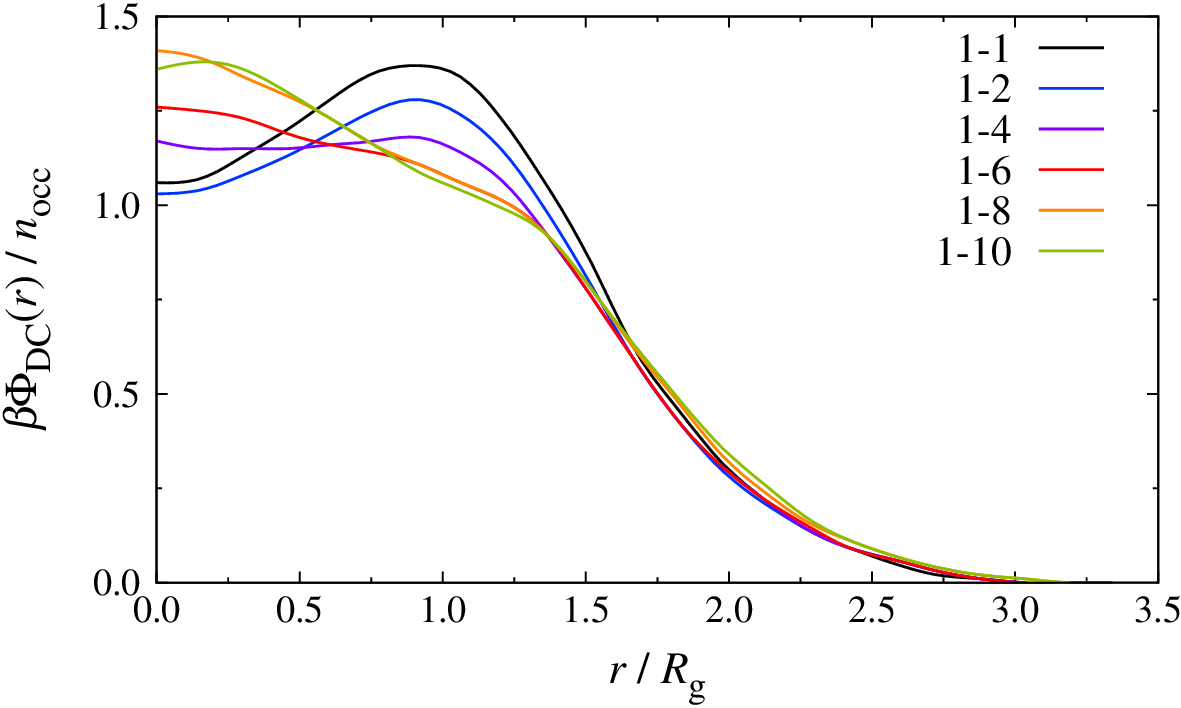} 
	\end{center}
	\caption{Effective potential $\Phi_{\rm DC}(r)$ between a single dendrimer and a 
	cluster of $n_{\rm occ}=2 - 10$ dendrimers (model I), computed in a MR simulation. 
	The $x$-axis has been scaled by the radius of gyration of the amphiphile in
	the zero density limit, $R_{\rm g}$, and the $y$-axis has been normalized by the 
	aggregation number $n_{\rm occ}$.}
	\label{fig:cluster_eff_pot1}
\end{figure}

In order to better understand the origin of this startling discrepancy, 
we computed the effective interaction between a single amphiphile and a 
cluster of these macromolecules for various occupation numbers $n_{\rm occ}$. 
Figure~\ref{fig:cluster_eff_pot1} shows $\Phi_{\rm DC}(r)/n_{\rm occ}$, 
{\it i.e.} the effective dendrimer-cluster potential normalized by the 
occupation number. These data show that $\Phi_{\rm DC}(r)$ became 
increasingly repulsive with increasing $n_{\rm occ}$, a trend which stemmed
from crowding effects in the cluster center. However, for sufficiently small 
$n_{\rm occ}$, $\Phi_{\rm DC}(r)$ was almost linearly additive with respect 
to $n_{\rm occ}$. 

\begin{figure}[htbp]
	\begin{center}
	\includegraphics[width=8.6cm]{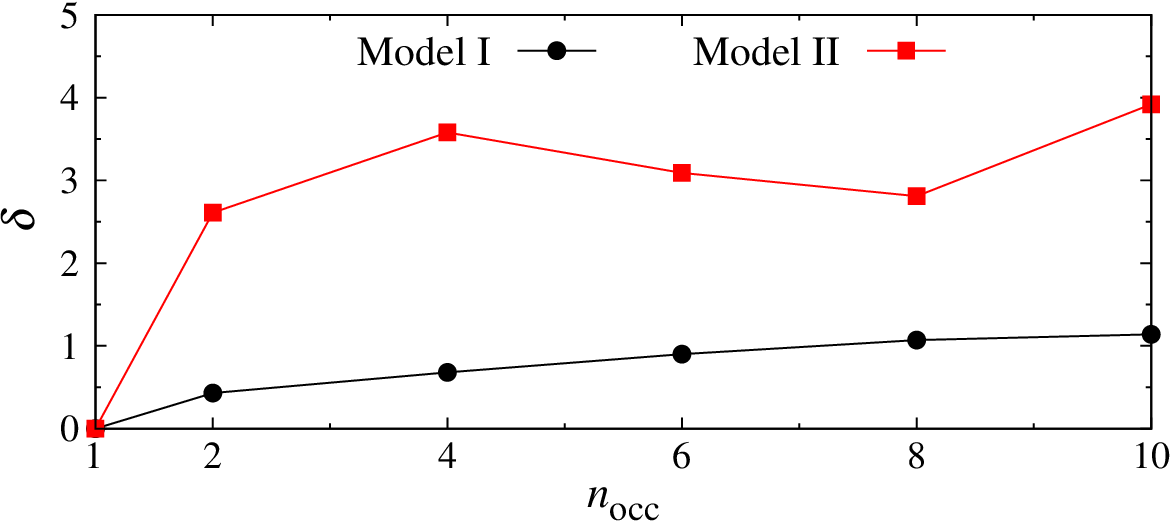}
	\caption{Deviation $\delta$ between the reduced dendrimer-cluster potential, 
	$\Phi_{\rm DC}/n_{\rm occ}$, and the effective dendrimer-dendrimer potential 
	in the zero-density limit, $\Phi_{\rm eff}^0$, as defined through Eq.~(\ref{eq:delta}). 
	Data shown for amphiphilic dendrimers of model I and II.}
	\label{fig:deviation}
	\end{center}
\end{figure}

We quantified the additivity of the potentials {\it via}:
\begin{align}
	\delta =  \int_0^{R_{\rm max}} \left|\Phi_{\rm eff}^0(r) - \Phi_{\rm DC}(r)/n_{\rm occ}\right|{\rm d}r .
	\label{eq:delta}
\end{align}
Figure~\ref{fig:deviation} shows $\delta$ as a function of $n_{\rm occ}$, 
and it is clearly visible that $\delta$ increased with $n_{\rm occ}$. 
Hence, the description using $\Phi_{\rm eff}^0(r)$ worsened with increasing 
$n_{\rm occ}$, resulting in erroneous structures in the CG description.

In order to understand the failure of $\Phi_{\rm eff}$, it is 
insightful to consider the cluster distribution $P(n_{\rm occ})$ 
shown in Figure~\ref{fig:cluster_amphi1}. It becomes immediately 
clear that the interactions in the system cannot be described 
through a {\it single} effective pair potential $\Phi_{\rm eff}$: 
at high densities, $P(n_{\rm occ})$ was rather broad, resulting 
in a large number of different dendrimer-cluster and cluster-cluster 
interactions. By mapping all these effective potentials onto a single 
$\Phi_{\rm eff}(r)$, we only preserve $\left\langle n_{\rm occ} \right\rangle$ 
but lose all information concerning the shape of $P(n_{\rm occ})$. This 
argument is corroborated by the fact that the $g(r)$ obtained from the 
MR and CG simulations agreed remarkably well for low density 
states $\rho \lesssim 0.52$, where the cluster distributions were 
rather narrow ({\it cf.} Figure~\ref{fig:cluster_amphi1}).

\subsection{Amphiphilic dendrimers - Model II}
To induce clustering of the amphiphiles at lower densities, 
considerable effort was put into tuning the interaction 
parameters of model I.\cite{lenz:prl:2012} In the revised model II, 
amphiphiles had a significantly more open core region, which was 
achieved by increasing the rest length of the central $g=0$ bonds. 
In addition, the range of the attraction between the core monomers 
was increased and thus acted well beyond the polymer's radius of gyration
($R_{\rm g} \approx 3.47 \pm 0.02$ for all investigated densities). 
These features successfully lowered the freezing density from 
$\rho^0_{\rm freeze} = 0.78$ (model I) to $\rho^0_{\rm freeze} = 0.141$. 
At this point we would like to mention, that an erroneous value of 
$\rho^0_{\rm freeze} = 0.281$ was reported originally in Ref.~\citenum{lenz:prl:2012} 
for the model II, due to a miscalculation of the corresponding effective 
potential $\Phi_{\rm eff}^0(r)$.

\begin{figure}[htbp]
	\begin{center}
	\includegraphics[width=8.6cm]{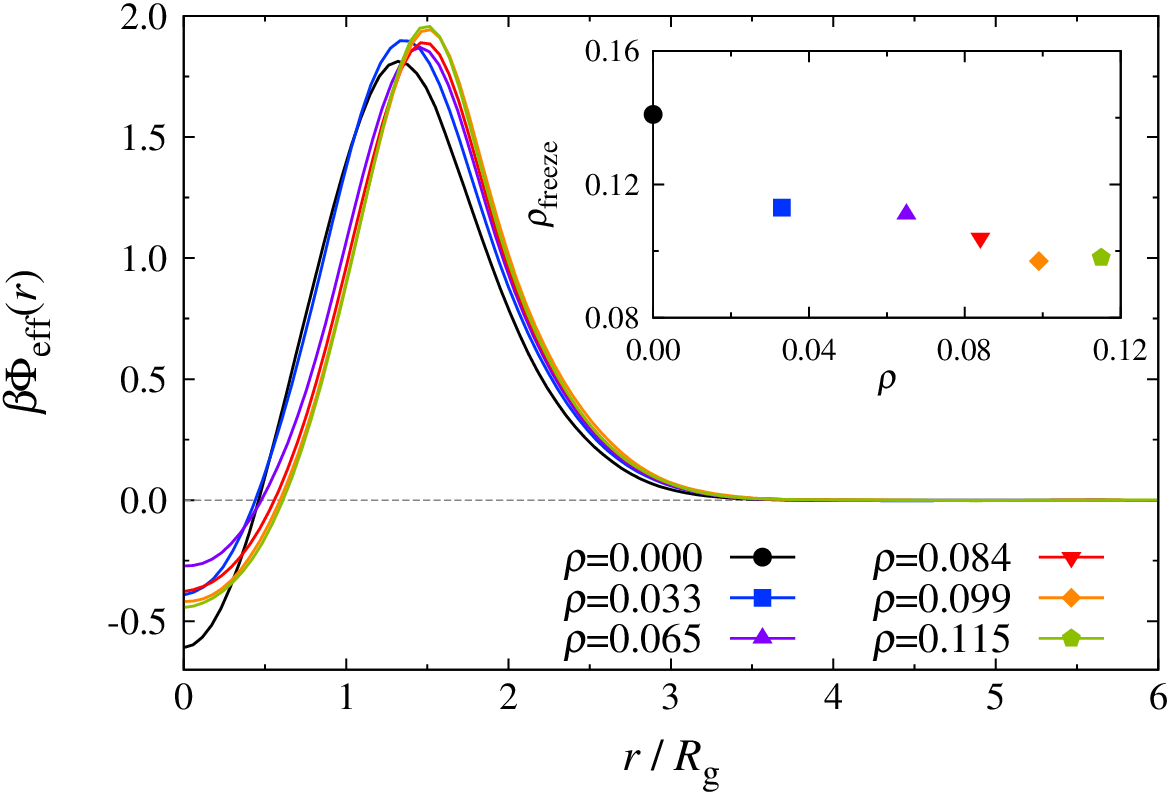} 
	\end{center}
	\caption{Same as Figure~\ref{fig:potential_amphi1}, but for model II.}
	\label{fig:potential_amphi2}
\end{figure}

We computed $\Phi_{\rm eff}(r)$ in the zero-density limit and at the 
reduced densities $\rho = 0.033$, $0.065$, $0.084$, $0.099$, $0.115$, 
where each system consisted of at least $N=1280$ dendrimers. We 
employed a basis of dimension $N_{\rm D}=20$ and a cutoff radius 
$R_{\rm max}=20.0$. The effective potentials $\Phi_{\rm eff}(r)$ 
are plotted in Figure~\ref{fig:potential_amphi2} for all 
investigated $\rho$ values. Only a very weak density-dependence of the 
effective potentials is discernible in this density regime. The inset of 
Figure~\ref{fig:potential_amphi2} shows the corresponding freezing 
densities $\rho_{\rm freeze}$, which were consistently lower than 
the $\rho_{\rm freeze}^0$ value.

\begin{figure}[htbp]
	\begin{center}
	\includegraphics[width=8.6cm]{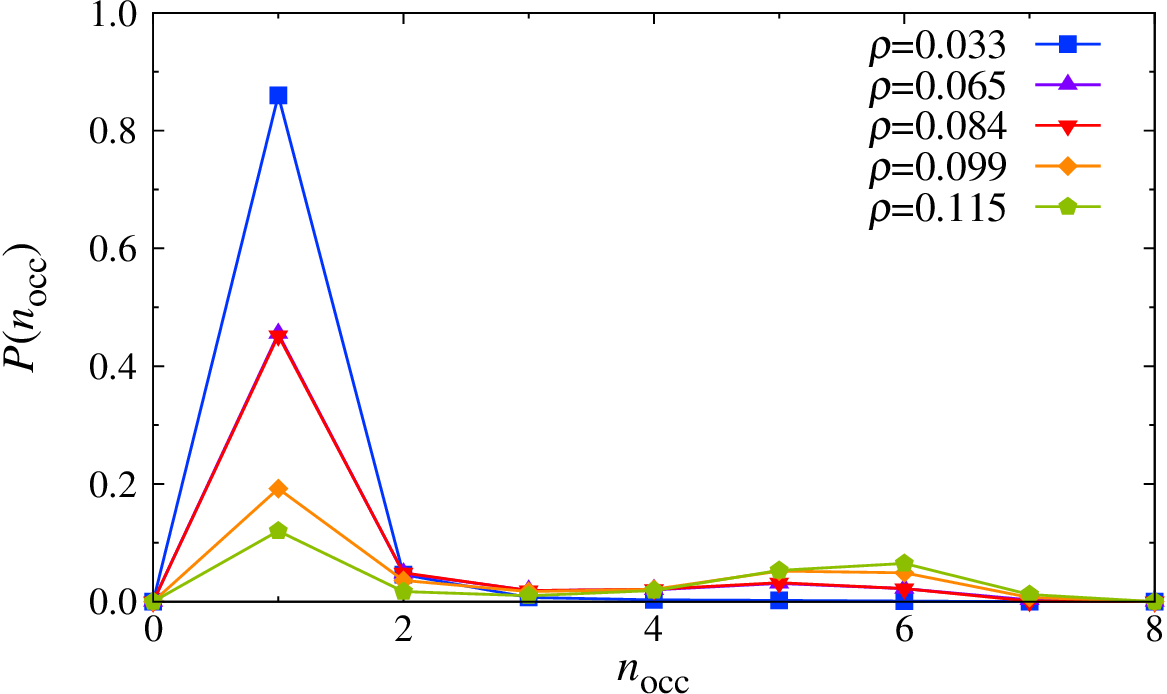} 
	\end{center}
	\caption{Same as Figure~\ref{fig:cluster_amphi1}, but for model II.}
	\label{fig:cluster_amphi2}
\end{figure}

Figure \ref{fig:cluster_amphi2} shows the cluster size distributions 
$P(n_{\rm occ})$ in the MR simulations, where we assigned dendrimers 
within a distance of $0.85\,R_{\rm g}$ to the same cluster. As $\rho$ 
was increased, the number of isolated amphiphiles decreased continuously 
and the dendrimers aggregated into clusters. At the same time, the 
local maximum at $n_{\rm occ} > 1$ became more pronounced and shifted 
towards larger $n_{\rm occ}$. The density at which clustering occurred 
was considerably lower compared to the model I case, resulting in 
lower occupation numbers $n_{\rm occ}$ ({\it cf.} Figure~\ref{fig:cluster_amphi1}). 

\begin{figure}[htbp]
	\begin{center}
	\includegraphics[width=8.6cm]{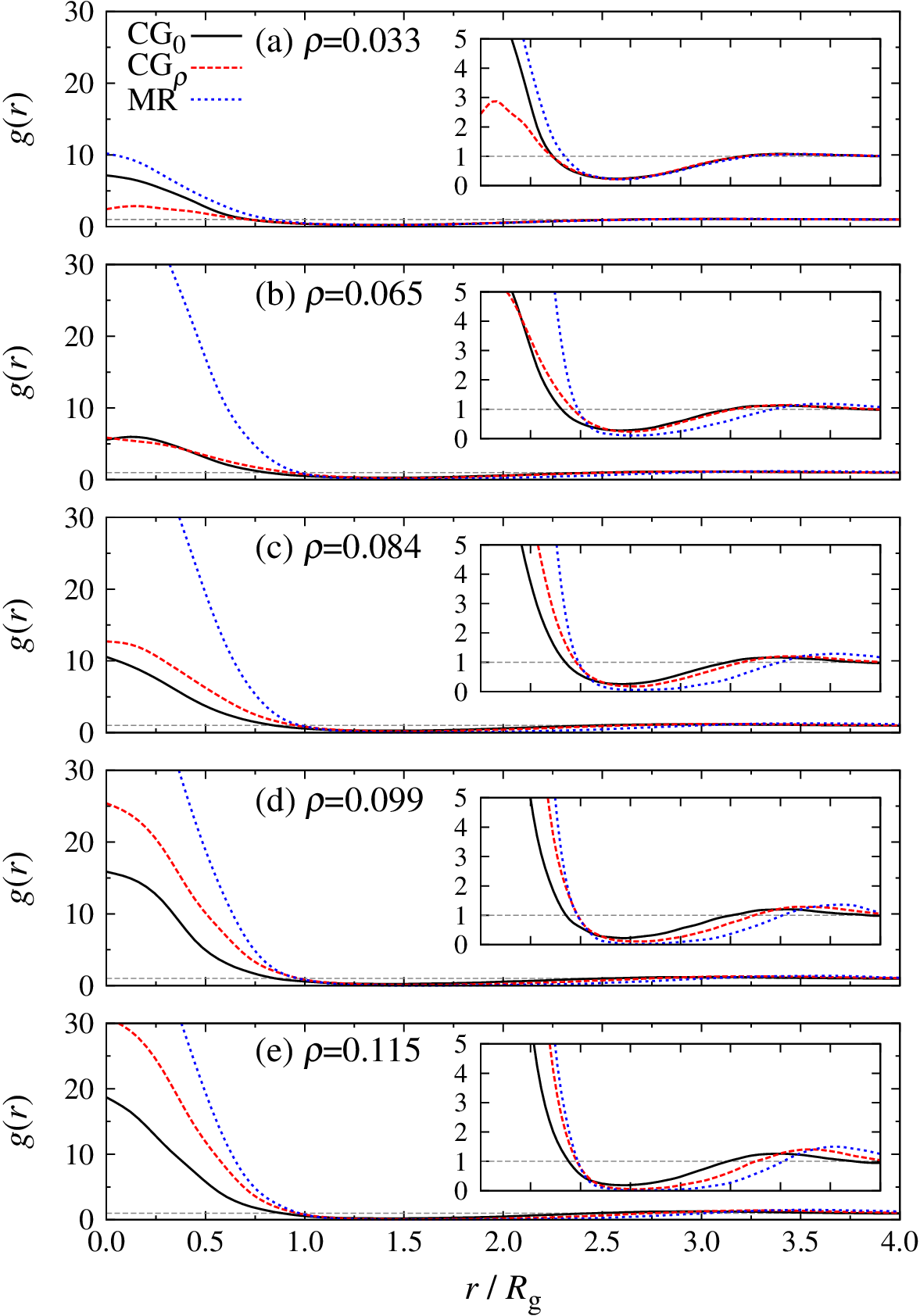} 
	\end{center}
	\caption{Same as Figure~\ref{fig:rdf_amphi1}, but for model II. The inset shows $g(r)$ with a zoomed in scale on the $y$-axis.}
	\label{fig:rdf_amphi2}
\end{figure}

At a first glance, the data presented in Figure~\ref{fig:potential_amphi2} 
and \ref{fig:cluster_amphi2} seem to suggest that a CG description using 
$\Phi_{\rm eff}(r)$ should produce good agreement with the MR simulations, 
since there was only a weak density-dependence on $\Phi_{\rm eff}(r)$ 
and the system had a rather narrow cluster distribution. However, 
both $\Phi_{\rm eff}^0(r)$ and $\Phi_{\rm eff}(r)$ failed to replicate the 
structures of the MR simulations, as evidenced by the radial distribution 
functions $g(r)$ plotted in Figure~\ref{fig:rdf_amphi2}. In fact, the 
difference between the MR and CG simulations was significantly more 
pronounced compared to the model I case investigated in 
Section~\ref{sec:model1} above. 

\begin{figure}[htbp]
	\begin{center}
	\includegraphics[width=8.6cm]{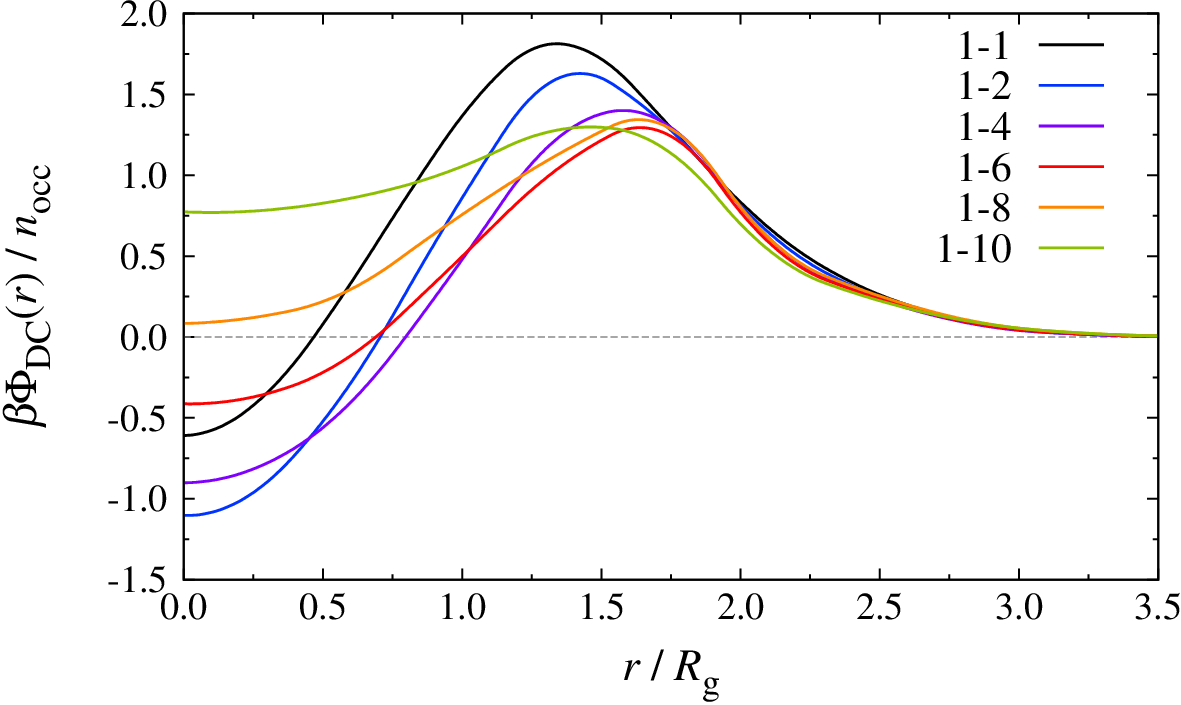} 
	\end{center}
	\caption{Same as Figure~\ref{fig:cluster_eff_pot1}, but for model II.}
	\label{fig:cluster_eff_pot2}
\end{figure}

To elucidate this surprising behavior, we computed $\Phi_{\rm DC}(r)$ 
for various occupation numbers $n_{\rm occ}$, analogous to our previous 
study of amphiphiles of model I (see Figure~\ref{fig:cluster_eff_pot1}). 
Figure~\ref{fig:cluster_eff_pot2} shows the resulting effective potentials
and we can see that the local minimum at $r=0$ first decreased but then 
increased rapidly as $n_{\rm occ}$ was increased. At the same time, the local 
maximum monotonically decreased and moved to slightly higher $r$ values. 

Figure~\ref{fig:deviation} shows the deviation of $\Phi_{\rm DC}/n_{\rm occ}$ 
[as defined by Eq.~(\ref{eq:delta})] with respect to the effective 
dendrimer-dendrimer potential in the zero-density limit, $\Phi_{\rm eff}^0(r)$. 
It is clearly visible that the deviations were significantly larger for the model 
II amphiphiles compared to the ones computed for model I, which explains the 
inferior agreement of the $g(r)$ shown in Figure~\ref{fig:rdf_amphi2}. In order 
to investigate a possible correlation between the conformation of the aggregated 
dendrimers and $\Phi_{\rm DC}(r)$, we measured the radial density distribution of 
the solvophobic core and solvophilic shell monomers. For the model II amphipiles, we 
observed a peculiar backfolding of the $g=1$ monomers into the core region, while the 
$g=2$ monomers formed the corona. In contrast, for the model I amphiphiles we observed 
a layered structure with the $g=0$ monomers in the core, the solvophobic $g=1$ monomers 
in the intermediate region, and the $g=2$ monomers in the shell. In general, we
observed that the core-shell structure became more pronounced as the cluster 
occupation number was increased. Furthermore, the conformation of the individual 
amphiphiles changed only marginally for the investigated values of $n_{\rm occ}$, 
suggesting that the pronounced variation of $\Phi_{\rm DC}(r)$ predominantly originated 
from excluded volume effects, which impeded the stacking of dendrimers.

\section{Conclusions}
\label{sec:conclusion_outlook}
We have calculated effective pair potentials in ultrasoft colloidal systems at finite densities, 
and have systematically compared the emerging structural properties in the microscopically resolved 
and coarse-grained simulations based on effective pair potentials. For non-clustering systems, we observed almost 
perfect agreement between the two representations, even at densities well above the
overlap density. However, a significant mismatch was observed for cluster-forming 
amphiphilic systems. This was surprising, since we employed effective pair potentials
which were computed from the microscopically resolved simulations at exactly the same 
temperatures and densities. 

The reason for this discrepancy is rooted in the heterogeneous density distributions
of the clustering systems. By taking the average over the whole system during the 
coarse-graining procedure, all information on these cluster distributions is lost. Such an
approach becomes problematic when the cluster sizes are not uniformly distributed, and 
the dendrimer-cluster and cluster-cluster interactions depend on the aggregation number 
of the partaking clusters. 

Hence, improved coarse-graining strategies are necessary for accurately describing 
clustering systems. For instance, it is conceivable to use a set of effective pair 
potentials, which correctly take into account the occupancy of the involved clusters. 
However, such a modification makes coarse-grained simulations considerably more expensive 
from a computational view, since they require a complete cluster analysis at each time step.
Therefore, further research is required to improve both accuracy and efficiency of such
coarse-grained simulations.

\begin{acknowledgments}
This work has been supported by the Marie Curie ITN-COMPLOIDS (Grant Agreement No. 234810), 
by the Austrian Science Fund (FWF) under Proj. Nos. P23910-N16 and F41 (SFB ViCoM), and by 
the WK ``Computational Materials Science'' of the Vienna University of Technology. Computing 
time on the Vienna Scientific Cluster (VSC) is gratefully acknowledged. A.N. acknowledges 
funding from the German Research Foundation (DFG) under the project number NI 1487/2-1.
\end{acknowledgments}

\end{document}